\documentclass[11pt,a4paper]{article}

\usepackage{jheppub}
\usepackage{appendix}
\usepackage{ mathrsfs }
\usepackage{graphicx}

\usepackage{dsfont}
\usepackage{physics}
\usepackage{wrapfig}
\usepackage{epstopdf}
\usepackage{mathtools}
\DeclarePairedDelimiter\ceil{\lceil}{\rceil}
\DeclarePairedDelimiter\floor{\lfloor}{\rfloor}
\usepackage{pifont}
\usepackage{amsmath}
\usepackage{amssymb}
\usepackage{xcolor}
\usepackage[T1]{fontenc}

\usepackage{verbatim}

\newcommand{\be}{\begin{equation}}
\newcommand{\ee}{\end{equation}}
\newcommand{\bea}{\begin{eqnarray}}
\newcommand{\eea}{\end{eqnarray}}

\newcommand{\gapp}{\mathrel{\raise.3ex\hbox{$>$}\mkern-14mu \lower0.6ex\hbox{$\sim$}}}
\newcommand{\lapp}{\mathrel{\raise.3ex\hbox{$<$}\mkern-14mu \lower0.6ex\hbox{$\sim$}}}

\newcommand{\HVEV}{\langle H\rangle}

\newcommand{\half}{\frac{1}{2}}

\def\bbox{{\,\lower0.9pt\vbox{\hrule \hbox{\vrule height 0.2 cm
\hskip 0.2 cm \vrule  height 0.2 cm}\hrule}\,}}

\title{$U(1)_Y\otimes \text{BRST}$ symmetry, of 
on-shell T-matrix elements and (1-$\phi$-{I}, 1-$A_{\mu}$-{R}) 
 Green's functions, determines the vacuum state of the Abelian Higgs Model from symmetry alone:
minimization of the scalar-sector effective potential 
is un-necessary
}

\author[a]{\"{O}zen\c{c} G\"{u}ng\"{o}r,} 
\author[a,b]{Bryan W. Lynn,} 
\author[a]{and Glenn D. Starkman}
\affiliation[a]{ISO/CERCA/Department of Physics, Case Western Reserve University, Cleveland, OH 44106-7079}
\affiliation[b]{Department of Physics and Astronomy, University College London, London WC1E 6BT, UK}

\emailAdd{oxg34@case.edu} 
\emailAdd{bryan.lynn@cern.ch}
\emailAdd{gds6@case.edu}

\abstract{
The  weak-scale Lorenz gauge $U(1)_Y$ Abelian Higgs Model (AHM)
is the simplest spontaneous-symmetry-breaking gauge theory: a scalar 
$\phi = \frac{1}{\sqrt 2}(H+i\pi)
\equiv \frac{1}{\sqrt 2}{\tilde H}e^{i{\tilde \pi}/\expval{H}}$; vector $A^\mu$; ghosts $\omega , {\bar \eta}$ decouple. T.W.B Kibble showed it has a Goldstone theorem:
$\tilde \pi$ ({\bf not} the linear pseudo-scalar $\pi$) is a massless 
derivatively coupled 
Nambu-Goldstone boson.

Global $U(1)_Y\otimes BRST$ symmetry \cite{LSS-3Proof} emerges, when it is realized that {\bf on-shell} T-matrix elements enjoy an extra $U(1)_Y$ global symmetry beyond the Lagrangian's BRST symmetry. The symmetries co-exist: $U(1)_Y$ generators $\delta_{U(1)_Y}$ commute 
with idempotent BRST generators $s$ and $\big[ \delta_{U(1)_Y}, s\big] {\cal L}_{AHM}=0$.
Two towers of  Ward Takahashi identities (WTI), which include all-loop-orders
 quantum corrections, emerge \cite{LSS-3Proof}:
a tower of relations among off-shell 1-$\phi$-I (but 1-$A_{\mu}$-{\bf{Reducible}}) Green's functions; another tower of Adler-zero WTI for on-shell T-matrix elements. 
The T-matrix's LSS theorem \cite{LSS-3Proof} forces tadpoles to automatically vanish
(equivalently $m_{\pi}^{2}=0$) {\bf by symmetry alone}. 

We show that, when the full symmetries of Lorenz gauge AHM are enforced in the scalar-sector effective potential, the vacuum state of the theory is specified/decided by symmetry alone. 
We use recursive WTI relations among 
Green's functions to include operators of $dimension\geq 1$.
We express the fully renormalized scalar-sector effective potential in a form which shows explicitly that, for small $\phi$-field values, the gauge-independent vacuum state of the theory $\expval{H}_{renormalized} = Z_{\phi}^{-1/2}\expval{H}_{bare}$ is determined by $U(1)_Y\otimes {BRST}$ symmetry alone, {\bf without minimizing the effective potential}. 

The extended-AHM (E-AHM) adds 
certain
($M_{\Phi}^2,M_{\psi}^2\sim M_{Heavy}^2 \gg \HVEV^2 \sim m_{Weak}^2$) $CP$-conserving heavy matter:
spin $S=0$ scalars $\Phi$ with $\big< \Phi\big>=0$; $S=\half$
anomaly-cancelling 
fermions $\psi$. {\bf By $U(1)_Y\otimes BRST$ symmetry alone}: the LSS theorem forces all heavy-particle relevant operators to vanish; $m_{\pi}^{2}=0$; tadpoles vanish; and the ground state is fully determined, with no need to minimize an E-AHM effective potential.
}


\notoc
\begin{document}
\maketitle
\flushbottom

\section{Introduction}

What are the symmetries of spontaneous symmetry breaking (SSB) Abelian Higgs Model (AHM) \emph{physics}? The symmetries of its \emph{Lagrangian} are well known \cite{Ramond2004}, but local gauge invariance is lost in the AHM Lagrangian, broken by gauge-fixing terms, and replaced with global BRST invariance \cite{BecchiRouetStora}. 
Nevertheless, Slavnov-Taylor identities \cite{JCTaylor1976} prove 
that the on-shell S-Matrix elements of \lq \lq \emph{physical states}" $A_{\mu},\, h,\, \pi$ 
(but not fermionic ghosts $\omega ,\, \bar{\eta}$) 
are independent of the usual anomaly-free $U(1)_{Y}$ 
gauge/local 
transformations 
even though these break the Lagrangian's BRST symmetry. 
B.W. Lynn and G.D. Starkman  observed, in collaboration with Raymond Stora  \cite{LSS-3Proof}, 
that they are therefore also independent 
of the anomaly-free $U(1)_{Y}$ \emph{global/rigid} transformations, 
resulting in \lq \lq new" global/rigid currents and appropriate $U(1)_{Y}$ Ward-Takahashi Identities (WTI).

In this paper we consider only the Goldstone (i.e. spontaneously broken) mode 
of the Abelian Higgs Model \cite{Higgs1964,Higgs1964-2,Englert1964,Guralnik1964} Lagrangian in the Lorenz gauge
\footnote{
\begin{equation}
\label{ahmlagrangian}
\begin{split}
\mathcal{L} &= \mathcal{L}_{Invariant} + \mathcal{L}_{GaugeFixing} + \mathcal{L}_{Ghost}, \\
\mathcal{L}_{Invariant} &= -\frac{1}{4} F_{\mu \nu}F^{\mu \nu} +  
\abs{D_{\mu}\phi}^{2}-\mu^{2}_{\phi}\phi^{\dagger}\phi - 
\lambda_{\phi}^{2}(\phi^{\dagger}\phi)^{2}, \\ 
\mathcal{L}_{GaugeFixing} &= - \lim_{\xi \rightarrow 0} \frac{1}{2 \xi}(\partial^{\mu}A_{\mu})^{2}, \\ 
\mathcal{L}_{Ghost} &= \bar{\eta}(-\partial^{2})\omega ,\\
\text{with} \quad 
\phi &\equiv \frac{1}{\sqrt{2}}(\expval{H}+h+i\pi), \, \, \, D_{\mu}\phi \equiv (\partial_{\mu} - ieY_{\phi}A_{\mu})\phi\,.
\end{split}
\end{equation}
Here $A_{\mu}$ is the $U(1)_{Y}$ gauge field,
$F_{\mu \nu}$ is the Abelian field-strength tensor for $A_{\mu}$,
$Y_{\phi}$ is the $U(1)_{Y}$ quantum number of $\phi$,
$\frac{1}{\sqrt{2}}\expval{H}$ is the expectation value of $\phi$ 
and $\omega,\bar{\eta}$ are the $U(1)_{Y}$ ghost and anti-ghost fields.
}
Naively, it has four apparently independent all-loop-orders renormalized parameters, 
$\lambda_{\phi}^{2}, \, \mu_{\phi}^{2}, \, e$ and $\expval{H}$. 
But the Lee-Stora-Symanzik (LSS) theorem \cite{LSS-3Proof}, i.e. the WTI
$m_{\pi}^{2} \equiv \mu_{\phi}^{2} + \lambda_{\phi}^{2}\expval{H}^{2}=0$, 
relates three of these parameters to one another by on-shell T-matrix $U(1)_Y$ symmetry.

To show that the symmetries of the theory encoded in the WTIs completely determine the vacuum state, we have to calculate the scalar-sector effective potential, while enforcing the symmetries. The scalar-sector effective potential involving only external scalars can be written as
\begin{equation}
\label{scalarveff}
V^{eff}_{\phi} = -\sum_{N,M=0}^{\infty}\frac{\Gamma_{N,M}}{N! M!}h^{N}\pi ^{M}
\end{equation}
The factorials account for the symmetry factors of the external states. 
The Green's functions $\Gamma_{N,M}$ have $N$ external $h$, and $M$ external $\pi$ legs, at zero momenta. 
They are one-scalar-particle-irreducible (1-$\phi$-I), but usually one-vector-particle-reducible (1-$A_{\mu}$-R). 
 \footnote{
If we were interested in the generators of vertex functions, 1-P-I (i.e. Irreducible in all types of fields) Green's functions  (see e.g. \cite{KrausSiboldAHM,Grassi1999} and classic textbooks \cite{ItzyksonZuber}) would be useful. 
But we are here instead interested in calculating the scalar-sector effective potential, i.e. in processes with external scalar particles only, so 1-$\phi$-I are much more useful.}
The set of 1-$\phi$-I Green's functions includes the more usual 1-P-I Green's functions, plus an infinity of certain other $1-P-Reducible$ graphs.

\subsection{Ward-Takahashi Identities (WTI): $U(1)_{Y}$ current is conserved for connected truncated Greens functions and T-matrix elements}

The scalar-sector effective potential must ultimately reflect 
the (quantum) symmetries of the scalar-sector of the theory. 
These are encoded in the Ward-Takahashi Identities (WTI).
We begin by focusing on the rigid/global $U(1)_{Y}$ current
\begin{eqnarray}
\label{AHMcurrent}
J^{\mu} &=& \pi \partial^{\mu}H - H \partial^{\mu}\pi - eA^{\mu}(\pi^{2}+H^{2})\,,
\\ \partial_\mu J^{\mu} &=& Hm_A \partial_\mu A^\mu \nonumber 
\\ m_A &=& Y_\phi e\HVEV \nonumber
\end{eqnarray}
embedded in the AHM local/gauge theory.
This current is conserved up to (ultra-soft) gauge-fixing terms, so that for Connected ($C$) time-ordered products with $N$ external $h$s and $M-1$ external $\pi$s, the gauge-fixing condition vanishes \cite{tHooft1971}
 \begin{eqnarray} 
 \label{GaugeCondition} 
\bra{0}T\big[(\partial_{\mu} A^\mu(z))h(x_{1})\ldots h(x_N)\pi(y_1)\ldots \pi(y_{M-1})\big]\ket{0}_{C} =0
 \end{eqnarray} 
 so that the current is effectively conserved
 \begin{eqnarray} 
 \label{CurrentConservation} 
\bra{0}T\big[(\partial_{\mu} J^\mu(z))h(x_{1})\ldots h(x_N)\pi(y_1)\ldots \pi(y_{M-1})\big]\ket{0}_{C} =0
 \end{eqnarray} 
For the spontaneously broken AHM in the Lorenz gauge, the Ward identities become (see \cite{LSS-3Proof} for details) \begin{equation} \label{masterwti} \begin{split} &- \expval{H}\partial^{\mu}_{z}\bra{0}T\big[(\partial_{\mu}\pi(z))h(x_{1})\ldots h(x_N)\pi(y_1)\ldots \pi(y_{M-1})\big]\ket{0}_{C} \\ &= i\sum_{m=1}^{M-1}\delta^{4}(z-y_{m})\bra{0}T\big[h(z)h(x_{1})\ldots h(x_N)\pi(y_1)\ldots \widehat{\pi(y_{m})} \ldots \pi(y_{M-1})\big]\ket{0}_{C} \\ &-i\sum_{n=1}^{N}\delta^{4}(z-x_{m})\bra{0}T\big[h(x_{1})\ldots \widehat{h(x_n)} \ldots h(x_N)\pi(z) \pi(y_1) \ldots \pi(y_{M-1})\big]\ket{0}_{C}\,, \end{split} \end{equation} where hatted fields are to be omitted. 
Ref. \cite{LSS-3Proof} derives 2 towers of WTIs, one for Green's functions and another for on-shell T-Matrix elements: they
exhaust the information content of \eqref{AHMcurrent},\eqref{CurrentConservation},\eqref{masterwti}. 

\subsection{$U(1)_Y\otimes BRST$ symmetry \cite{LSS-3Proof}}
In G. 't Hooft's $R_\xi$ gauges, gauge fixing and DeWitt-Fadeev-Popov ghost terms 
\cite{DeWitt1967,Fadeev1967}
are written in terms of a Nakanishi-Lautrup field $b$
\cite{Nakanashi1966,Lautrup1967}, 
and the SSB vector mass $m_A=eY_\phi\HVEV =-e\HVEV >0$.
\bea
\label{tHooftGaugeFixing}
\mathcal{L}_{AHM}^{R_\xi}=\mathcal{L}_{AHM}^{Invariant}&+&\mathcal{L}_{AHM}^{GaugeFix;R_\xi}+ \mathcal{L}_{AHM}^{Ghost;R_\xi} \nonumber \\
\mathcal{L}_{AHM}^{GaugeFix;R_\xi}&+& \mathcal{L}_{AHM}^{Ghost;R_\xi} \nonumber \\
&=&s\Big[{\bar \eta} \Big(F_A + \half \xi b \Big)\Big] \nonumber \\
F_A &=&\partial_\mu A^\mu +\xi m_A \pi 
\end{eqnarray}
with global BRST transformations \cite{BecchiRouetStora,Tyutin1975,Tyutin1976,Nakanashi1966,Lautrup1967,Weinberg1995} $s$, the Lagrangian 
is BRST invariant: 
\bea
s \mathcal{L}_{AHM}^{R_\xi} = 0
\eea

Reference \cite{LSS-3Proof} defines the properties of the various fields under anomaly-free un-deformed rigid/global $U(1)_Y$ transformation by a constant $\Omega$, and we discover that the $R_\xi$-gauge Lagrangian 
is not invariant 
under such $U(1)_Y$ transformations
\bea
\label{U(1)TransformationLagrangianSummary}
\delta_{U(1)_Y} \mathcal{L}_{AHM}^{R_\xi} &\neq& 0
\eea
Still, 
the actions of the {\bf BRST transformations 
and the $U(1)_Y$ transformation 
commute} on all fields (for the definition of these transformations, see \cite{LSS-3Proof} eq.s (7)-(10)).  
Thus, with the nilpotent property $s^2=0$ applied in \eqref{tHooftGaugeFixing}
\bea
\label{U(1)BRSTCommutatorAHM}
\Big[ \delta_{U(1)_Y}, s \Big] \mathcal{L}_{AHM}^{R_\xi} &=& 0\,,
\eea 
and the two separate global $BRST$ and on-shell $U(1)_Y$ symmetries can therefore co-exist in AHM  physics.

\subsection{$1-\phi-I$ (but $1-A_\mu-R)$ Greens functions}
The WTIs of the Goldstone-mode AHM 
for connected truncated one-scalar-particle-irreducible (1-$\phi$-I) Green's functions,
with $N$ renormalized $h$ fields with momenta $p_{i}$ and coordinate label $x_{n}$, 
with $M-1$ renormalized $\pi$ fields with momenta $q_{j}$ and coordinate label $y_{m}$, 
and with 1 renormalized  zero-momentum $\pi$ with coordinate label $z$, are \cite{LSS-3Proof}
\begin{eqnarray}
\label{wtiahm}
&\expval{H}&\Gamma_{N,M}(p_{1},\ldots ,p_{N};0,q_{1},\ldots ,q_{M-1}) \\ 
&=& \sum_{m=1}^{M-1}\Gamma_{N+1,M-2}
(q_{m},p_{1},\ldots ,p_{N};q_{1},\ldots,\widehat{q_{m}},\ldots ,q_{M-1}) \nonumber \\ 
&-& \sum_{n=1}^{N}\Gamma_{N-1,M}
(p_{1},\ldots ,\widehat{p_{n}},\ldots, p_{N};p_{n},q_{1},\ldots ,q_{M-1})\nonumber \,,
\end{eqnarray} 
where hatted momenta again represent omitted fields. 
As we are interested in the effective potential, a more convenient form of the WTIs at {\bf zero external momenta} is
\begin{equation}
\label{mastereq}
\begin{split}
\Gamma_{N,M} &=\frac{\expval{H}}{M+1}\Gamma_{N-1,M+2}+\frac{N-1}{M+1}\Gamma_{N-2,M+2}, \\ \Gamma_{N,M} &\equiv \Gamma_{N,M}(0,\ldots,0;0,\ldots,0).
\end{split}
\end{equation}
This encodes symmetries of the theory and can be used to put the expression \eqref{scalarveff} for the effective potential in a form that makes it explicit that the symmetries completely determine the vacuum state.

\subsection{$1-\phi-R$ ($1-A_\mu-R$) on-shell T-Matrix elements}
The  WTIs (\ref{mastereq}) exhaust the symmetries (including BRST) of 
the connected truncated zero-external-momentum Green's functions, 
but the T-matrix still has additional symmetry 
due to properties of the physical states \cite{LSS-3Proof}.

On-shell processes T-matrix elements are governed by an additional $U(1)_Y$ symmetry
This is reflected in the Adler self-consistency relations (i.e. ``Adler zeros") \cite{Adler1965},
written in terms of connected truncated on-shell 
one-scalar-particle-\emph{reducible (1-$\phi$-R)} T-matrix elements, 
with $N$ renormalized $h$ fields with momenta $p_{i}$ and coordinate label $x_{n}$,
$M-1$ renormalized $\pi$ fields with momenta $q_{j}$ and coordinate label $y_{m}$, 
and 1 renormalized zero-momentum $\pi$ with coordinate label $z$. 
For the AHM these are \cite{LSS-3Proof}
\begin{equation}
\label{adlertmatrix}
0 = \expval{H}T_{N,M}(p_{1},\ldots ,p_{N};0,q_{1},\ldots ,q_{M-1})(2\pi)^{4}\Big(\sum_{n=1}^{N}p_{n} + \sum_{m=1}^{M}q_{m}\Big)_{p_{n}^{2}=m_{h}^{2},q_{m}^{2}=0}.
\end{equation}
The $N=0,\, M=2$ case:
\begin{equation}
\label{adlerscT}
\expval{H}T_{0,2}(;0,0) = 0
\end{equation}
ensures that  $\pi$ remains massless to all orders in quantum loops.
 \footnote{
	This appears remarkably like the Goldstone Theorem, 
	which demands that  the Nambu Goldstone Boson (NGB)
	of a spontaneously broken global symmetry be purely derivatively coupled
	and so have zero mass to all orders in quantum loops.
	However, $\pi$ is not that derivatively coupled field. 
	The actual NGB, $\tilde\pi$ is found 
	by transforming $\phi$ to the Kibble representation:
	$\phi=\frac{1}{{\sqrt 2}} {\tilde H}e^{i {\tilde \pi}/\expval{H}}$,
	a transformation that can be done whenever $\langle H\rangle\neq0$, 
	i.e.  in the spontaneously broken theory.
	$\tilde\pi$ {\em is} derivatively coupled and hence massless, $m_{\tilde\pi}=0$,
	as per the Goldstone Theorem,
	despite that $\tilde\pi$ decouples from 
	the other propagating degrees of freedom
	with appropriate re-definition of the vector field. 
	For a discussion of the existence of the Goldstone Theorem 
	in the spontaneously broken $U(1)_{Y}$ gauge theory in Lorenz gauge see \cite{Kibble1967}.
	In the meantime we see that the vanishing of $m_\pi^2$, i.e. of $\Gamma_{0,2}$,
	is an independent piece of information from the Goldstone Theorem in the gauge theory.
	We refer to this as the B.W. Lee, R. Stora and K. Symanzik (LSS) Theorem, 
	since it is one of the on-shell T-Matrix WTI obtained by those three in global sigma models (with or without PCAC). 
	The distinctness of the LSS theorem from the Goldstone Theorem 
	in spontaneously broken local gauge theories in the presence of global BRST symmetry, 
	was first appreciated with Raymond Stora in \cite{LSS-3Proof}.}
Eqn. (\ref{adlerscT}) is the Lee-Stora-Symanzik (LSS) theorem, relating $m_{\pi}^{2}$ to an on-shell T-matrix element. This statement is true for all orders in loops written in this way as a T-matrix identity.
Observing that 
\begin{equation}
\begin{split}
\label{2point}
T_{0,2}(;0,0) &= \Gamma_{0,2} \\ &= -m^{2}_{\pi},
\end{split}
\end{equation} 
the Adler self-consistency condition \eqref{adlerscT} for $N=0,M=2$ 
also demands that
\begin{equation}
\label{adlersc}
\expval{H}\Gamma_{0,2}= -\expval{H}m_{\pi}^{2}=0.
\end{equation}

A crucial effect of the LSS theorem, together with the $N=1, M=0$ WTI in \eqref{mastereq} is to automatically eliminate the tadpoles in the scalar-sector effective potential \eqref{scalarveff} so no explicit tadpole renormalization is necessary. It can also be seen that, applying \eqref{mastereq} recursively in the presence of CP conservation, the Adler self-consistency relations for the AHM also enforce that Green's functions with an odd number of external $\pi$ legs vanish, 
which we use below in the derivation of the scalar-sector effective potential 
of equation \eqref{scalarveff}. 

Finally, the renormalization procedure is expressed by fixing  the quartic coupling constant 
to be related to the 4-point 1-$\phi$-I Green's function at zero external momenta
\begin{equation}
\label{renormalization}
\lambda_{\phi}^{2}= -\frac{1}{6}\Gamma_{0,4}\,.
\end{equation}

\subsection{Effective Lagrangian for $dimension \leq 4$ operators}
When the renormalization condition in \eqref{renormalization} 
and the Adler self-consistency relations are enforced, 
the effective scalar-sector Lagrangian for the AHM becomes \cite{LSS-3Proof}, 
for terms of dimension $\leq4$
\begin{equation}
\label{leffAHM}
\begin{split}
\mathcal{L}_{\phi}^{eff} &= 
\abs{D_{\mu}\phi}^{2} - V_{\phi}^{eff} \\
V_{\phi}^{eff} &= \lambda_{\phi}^{2}\zeta^{2} \\
\zeta &= \phi^{\dagger}\phi - \frac{\expval{H}^{2}}{2} = \frac{h^2+\pi^2}{2} +\HVEV h
\end{split}
\end{equation}

In \cite{LSS-3Proof}, the authors focused on terms with dimension less than or equal to four. 
Below, we will focus on terms with dimensions greater than 4 
and enforce the WTIs in \eqref{mastereq} 
to greatly simplify the full effective scalar-sector potential.

Written suggestively, the LSS theorem in Goldstone mode is
\begin{eqnarray}
\label{Vmin}
\expval{H} &\neq& 0; \quad\quad \mu_{\phi}^{2}+\lambda_{\phi}^{2}\expval{H}^{2} = 0
\end{eqnarray}

It is instructive to compare the results of the LSS theorem \eqref{adlerscT} with the current literature, which  agrees with \eqref{Vmin}, but regards it as arising from the minimization of the effective potential.
According to that view, after renormalization, all ultraviolet quadratic divergences and relevant-opeerator contributions to $\expval{H}$, including those from very heavy particle representations added to the AHM in the E-AHM,  are regarded as cancelled against a bare counter-term $\delta \mu_{\phi}^{2}$. {\bf In the absence of the LSS theorem}, no symmetry protects \eqref{Vmin}. 

In stark contrast, in this work, the LSS theorem (i.e. the symmetriy of on-shell T-Matrix elements and their Adler zeros) protects \eqref{Vmin} exactly. We will never minimize a potential to calculate $\expval{H}$, and we will never explicitly renormalize tadpole contributions. Enforcing the LSS theorem makes the tadpoles vanish {\bf by symmetry alone} and enforcing the WTIs on the scalar-sector effective potential will decide the true vacuum of the theory.

\section{The all-loop-orders, $\text{dimension} \geq 1$ operators, scalar-sector effective potential in Lorenz gauge AHM}
\subsection{Greens functions at zero external momenta}

The Appendix shows that equation \eqref{mastereq} can be solved recursively 
to express $\Gamma_{N,M}$ as a linear combination of $\Gamma_{0,m}$\,, 
thus expressing all relevant Green's functions in terms of Green's functions 
with no external $h$ legs: 
\begin{equation}
\label{recursive}
\Gamma_{N,M} = \sum_{k=0}^{\floor{\frac{N}{2}}}\frac{(M-1)!! N!}{(M+2N-2k-1)!!} \frac{\expval{H}^{N-2k}}{(2k)!!(N-2k)!}\Gamma_{0,M+2(N-k)}
\end{equation}
where $\floor{x}$ is defined as
$\floor{x} \equiv  \text{max}\{x \in \mathbb{Z} \, \vert \, m \leq x \}$. 
Using eq. \eqref{recursive} in \eqref{scalarveff} and the LSS theorem \eqref{adlersc}, (details can be found in the Appendix) the effective potential for the scalar sector becomes
\begin{equation}
\label{veffxi}
V_{\phi}^{eff} =-\sum_{n=0}^{\infty}\frac{1}{(2n){!}}\Gamma_{0,2n}\Big(h^{2}+\pi^{2}+2\expval{H}h\Big)^{n}.
\end{equation}
\\
The Green's functions $\Gamma_{0,2n}$ are renormalized to all orders in quantum loops,
and 1-$\phi$-I but not 1-$A_\mu$-I.  
They contain the symmetries encoded in the WTIs. 
For arbitrary $n$,
$\Gamma_{0,2n}$ can be calculated to order $k$ in loops 
to generate the correct renormalized effective potential up to that loop order.

The scalar-sector effective potential $V^{eff}_{\phi}$ in \eqref{veffxi}
respects the symmetries of the theory through the WTIs. 
It is built out of connected truncated 1-$\phi$-I (but not 1-$A_{\mu}$-I) Green's functions. 
The renormalization condition is that 
the quartic coupling constant is related to the 1-$\phi$-I Green's function 
with four external $\pi$ legs at zero external momenta as in \eqref{renormalization}.
The Green's function version of the LSS Theorem \eqref{adlersc}, 
and the renormalization condition \eqref{renormalization}, 
must therefore be imposed on \eqref{veffxi}. 

The $n=0$ term is composed of disconnected diagrams, 
which are excluded in the T-Matrix, having been absorbed into an overall phase \cite{Bjorken1965} it shares with the vacuum. 
\begin{equation}
\label{Greensveffxi}
V_{\phi}^{eff} =-\sum_{n=1}^{\infty}\frac{1}{(2n){!}}\Gamma_{0,2n}\Big(h^{2}+\pi^{2}+2\expval{H}h\Big)^{n}.
\end{equation}

\subsection{Lee-Stora-Symanzik theorem}
The \emph{physics} of $V^{eff}_{\phi}$ must also obey the further symmetries of the on-shell T-Matrix: i.e. connectedness and the LSS Theorem \eqref{adlerscT}.
\begin{quote}
{\it ``Whether you like it or not, 
you have to include in the Lagrangian all possible terms consistent with locality and power counting, 
unless otherwise constrained by Ward identities."}
Kurt Symanzik, in a private letter to Raymond Stora \cite{SymanzikPC} 
\end{quote}
In strict obedience to K. Symanzik's edict, 
we now further constrain the allowed terms in the $\phi$-sector effective potential,
using  those $U(1)_Y$ Ward-Takahashi identities 
that govern 1-$\phi$-R T-Matrix elements $T_{N,M}$.

The $n=1$ term vanishes because of the LSS Theorem \eqref{adlersc}: 
Green's functions and T-Matrix  elements with zero external $h$ particles 
and two external $\pi$ particles vanish. The LSS theorem, as explained in the previous section makes the tadpole contributions vanish as well. 
We are left with 
\begin{equation}
\begin{split}
V^{eff}_{\phi} = 
- \sum_{n=2}^{\infty}\frac{1}{(2n){!}}\Gamma_{0,2n}\Big(h^{2}+\pi^{2}+2\expval{H}h\Big)^{n}\,,
\end{split}
\end{equation}
where $n=2$ is fixed by the renormalization condition in \eqref{renormalization}. 
Re-expressing $V_{\phi}^{eff}$ in terms of the variable $\zeta$ in (\ref{leffAHM}), 
the scalar-sector effective Lagrangian is
\begin{equation}
\label{leffAHMLorenz}
\begin{split}
\mathcal{L}_{\phi}^{eff} &= 
\abs{D_{\mu}\phi}^{2} - V_{\phi}^{eff} \\
V_{\phi}^{eff} &= \lambda_{\phi}^{2}\zeta^{2} - \sum_{n=3}^{\infty}\frac{2^{n}}{(2n){!}}\Gamma_{0,2n}\zeta^{n} \,.
\end{split}
\end{equation}
This governs low-energy scalar-sector physics in Lorenz gauge.

We find it instructive to re-state what this expansion achieves; by using the symmetries of the physical states of the theory encoded in the WTIs and the on-shell T-matrix elements and expressing the scalar sector effective potential in terms of connected, $1-\phi -I$ Green's functions it is explicitly seen that the vacuum of the theory is decided by the symmetries alone of the theory, no minimization is necessary to calculate the value of $\expval{H}$ or to decide the correct vacuum of the theory.


\smallskip
\smallskip
\section{The gauge-independent vacuum state of the SSB AHM theory is determined by symmetry alone} 

To see that the theory has the usual Nambu-Goldstone boson (NGB) $\tilde{\pi}$, 
we transform to the  Kibble representation 
$\phi = \frac{1}{\sqrt{2}}\tilde{H}e^{i\frac{\tilde{\pi}}{\expval{H}}}$ with $Y_{\phi} = -1$,
$\tilde{H}=\expval{H}+\tilde{h}$, and re-write (\ref{leffAHMLorenz}) 
After the $\tilde \pi$ NGB decouples, 
the effective Lagrangian which governs low-energy scalar-sector physics becomes
\bea
\label{leffAHMLorenzB}
\mathcal{L}^{eff}_{\phi} &=& \frac{1}{2}\Big\vert (\partial_{\mu}+ieB_\mu){\tilde H}\Big\vert^{2} 
-V_{\phi}^{eff} \nonumber \\
V_{\phi}^{eff} &=& \frac{\lambda_{\phi}^{2}}{4}\Big( {\tilde H}^2-\HVEV^2\Big)^{2} + \sum_{n=3}^{\infty}V_{2n}\Big( {\tilde H}^2-\HVEV^2\Big)^{n} \nonumber\\
\zeta &=& \phi^{\dagger}\phi - \frac{\expval{H}^{2}}{2} = \half\Big( {\tilde H}^2-\HVEV^2\Big) \nonumber \\
{\tilde H} &=& {\tilde h} +\HVEV; \qquad \big<{\tilde h}\big>=0.
\eea
where we have defined the physical/observable gauge field 
\begin{equation}
B_{\mu} = A_{\mu} + \frac{1}{e\expval{H}}\partial_\mu \tilde{\pi},
\end{equation}
the NGB $\tilde{\pi}$ has been ``eaten" by the vector field 
and the physical external states consist only of the 
Brout-Englert-Higgs boson $\tilde{h}$ with mass
$m^{2}_{\tilde{h}} = 2\lambda^{2}_{\phi}\expval{H}^2$ 
and the massive gauge boson $B_{\mu}$ 
with mass $m^{2}_{B}=e^2\expval{H}^2$. $V_{2n}\equiv-\Gamma_{0,2n}$ in the Kibble representation.

The extrema of the potential, 0 and $\expval{H}$, are proved gauge-independent by Nielson \cite{Nielsen1975}.
$\lambda_{\phi}^{2}$, $V_{2n}$ and (\ref{leffAHMLorenzB}) are proved gauge-independent by S.-H. Henry Tye and Y. Vtorov-Karevsky \cite{Tye1996}, when calculated in the Kibble representation. 

 If $V^{eff}_\phi$ is bounded from below, the minimum of the effective potential \eqref{leffAHMLorenzB} {\bf also} occurs at $\langle \tilde{H} \rangle = \expval{H}$. But we never minimized the effective potential.
Instead, we imposed on \eqref{leffAHMLorenzB} the Green's function WTIs and the LSS theorem. As a result, the theory has picked the vacuum itself from symmetry alone, i.e. the vacuum in which $\langle \tilde{H} \rangle$ takes the vacuum expectation value $\expval{H}$. 
As long as the coefficients of the higher powers in the effective potential in \eqref{leffAHMLorenzB} are positive or negative but systematically smaller in magnitude, the scalar-sector effective potential, if bounded from below and for small $\phi$ field values, will not generate new extrema, and the true vacuum of the theory will be at $\langle \tilde{H} \rangle = \expval{H}$. The numerical value of $\frac{\HVEV}{GeV}$ is an input parameter, to be taken from experiment.

\section{Addition of very heavy $U(1)_Y$ particles in loops: all relevant operators again vanish due to the LSS theorem; the vacuum is determined by $U(1)_Y\otimes BRST$ symmetry alone.}
The extended-AHM (E-AHM) adds 
certain
($M_{\Phi}^2,M_{\psi}^2\sim M_{Heavy}^2 \gg \HVEV^2 \sim m_{Weak}^2$) $CP$-conserving heavy matter:
spin $S=0$ scalars $\Phi$ with $\big< \Phi\big>=0$; $S=\half$
anomaly-cancelling 
fermions $\psi$. $1-\Phi -R$ graphs must now be included in Green's functions.
Results analogous to the AHM \eqref{leffAHMLorenzB} hold for the E-AHM to all-loop-orders for $dimension \geq 1$ operators (see \cite{LSS-3Proof} for $dimension\leq 4$ operators). {\bf By $U(1)_Y\otimes BRST$ symmetry alone}: the LSS theorem 
\bea
\big[ m_{\pi}^{2}\big]_{E-AHM}=\big[ \mu_\phi^2+\lambda^2\HVEV^2 \big]_{E-AHM}=0
\eea
forces all heavy-particle relevant operators to vanish; tadpoles vanish. The $\phi$-sector effective potential is
\begin{equation}
\label{E-AHMVeff}
V^{eff}_{\phi,\text{E-AHM}} = -\sum_{n=2}^{\infty}\frac{1}{(2n)!}\Gamma^{\text{E-AHM}}_{0,2n}\left(\tilde{H}^2-\expval{H}\right)^n
\end{equation}

and the SSB ground state ${\tilde H}^2=\HVEV^2$ is fully determined, with no need to minimize an E-AHM effective potential. The numerical value of $\frac{\HVEV}{GeV}$ is again an input parameter, to be taken from experiment.

\section{Conclusions}
We have solved the WTIs of the Goldstone mode of the Abelian Higgs Model recursively 
to express 1-$\phi$-I connected truncated Green's functions at zero momentum 
in Lorenz gauge, in terms of such Green's functions at
zero momentum with no external $h$ legs.
We  enforced the Lee-Stora-Symanzik (LSS) theorem on the solution, 
thus ensuring that the pseudoscalar $\pi$ remains massless 
to all orders in quantum loops, 
and forcing Green's functions with an odd number of external $\pi$ legs to vanish. The LSS theorem also makes the tadpole contributions vanish so we don't have to explicitly renormalize tadpole contributions.
The renormalization condition is 
expressed through the relation of the quartic coupling constant 
to the  zero-momentum Green's function with four external $\pi$ 
and zero external $h$ legs (equation \eqref{renormalization}). 
Together with the Green's functions, the Adler self-consistency relations and the LSS Theorem,
the recursive solution to the WTIs includes all the symmetries of the theory.


We have shown that imposing the full symmetries of the theory (the WTIs and the LSS theorem) on the effective potential ensures that the vacuum of the theory is where $\langle \tilde{H} \rangle = \expval{H}$. We have never minimized the effective potential to reach that conclusion, the theory picked the correct vacuum after the symmetries are imposed on the potential itself. By the use of the LSS theorem, the tadpole contributions vanish as well, saving us from doing explicit tadpole renormalization. The numerical value of $\frac{\HVEV}{GeV}$ is an input parameter, to be taken from experiment.

The 2 towers of WTIs derived in \cite{LSS-3Proof} are in the Lorenz gauge where,
in the Kibble representation, the $\tilde{\pi}$ field is \lq \lq eaten" by the observable vector field $B_{\mu}$ and decouples from the observable particle spectrum of $\tilde{h}, B_{\mu}$. The vacuum of the AHM, expressed in the Kibble representation, is gauge independent as proven by \cite{Tye1996}.

Results analogous to the AHM hold for the $CP$-conserving E-AHM to all-loop-orders for $dimension \geq 1$ operators. The numerical value of $\frac{\HVEV}{GeV}$ is again an input parameter, to be taken from experiment.

The arguments used and the solution provided can be extended to the scalar sector of the $CP$-conserving Standard model ($SM_{CP}$) for all-ElectroWeak and QCD-loop-orders $dimension \leq 4$ operators \cite{LSS-4Proof}, where $1-W_\mu^\pm -R$,  $1-Z_\mu -R$, $1-A_\mu (photon) -R$ and $1-G^a_\mu (gluon) -R$ graphs must be included in $1-Scalar-I$ Green's functions. The driving symmetry is $SU(2)_L\otimes BRST$. We expect to be able to extend our results to $dimension \geq 1$ operators in its scalar-sector.

The extended $CP$-conserving SM ($E-SM_{CP}$) adds 
certain $CP$-conserving
($M_{\Phi}^2,M_{\psi}^2\sim M_{Heavy}^2 \gg \HVEV^2 \sim m_{Weak}^2$) heavy matter:
spin $S=0$ scalars $\Phi$ with $\big< \Phi\big>=0$; $S=\half$
anomaly-cancelling 
fermions $\psi$. $1-\Phi -R$ graphs must also be included in $1-scalar-I$ Green's functions.
We have shown results analogous to those here  for all-loop-order $dimension \leq 4$ operators \cite{LSS-5Proof}.
We expect that analogous all-loop-order results also hold for the $E-SM_{CP}$ for $dimension \geq 1$ operators, so that {\bf by $SU(2)_L\otimes BRST$ symmetry alone}: the LSS theorem forces all heavy-particle relevant operators to vanish; $m_{\pi}^{2}=0$ for weak-isospin $\vec \pi$; tadpoles vanish; and the SSB ground state is fully determined/specified, with no need to minimize an $E-SM_{CP}$ effective potential.

\acknowledgments
OG and GDS are partially supported by  grant DOE-SC0009946 from the US Department of Energy.
BWL thanks Jon Butterworth and University College London 
for support as a UCL Honorary Senior Research Associate. 

\bibliography{references.bbl}
\bibliographystyle{unsrt}

\appendix
\section{Solving the Ward-Takahashi Identities and the Scalar-Sector Effective Potential}
\label{AllOrdersScalarPotential}
Equation \eqref{mastereq} is a recursion relation among the 1-$\phi$-I connected truncated  Green's functions of the theory. It relates a Green's function with $N$ external $h$ legs and $M$ external $\pi$ legs to two Green's functions, both with fewer external $h$ legs -- one with $N-1$ external $h$ and $M+2$ external $\pi$ legs, and the other with $N-2$ external $h$ and $M+2$ external $\pi$ legs. This relation can then be applied again and again until one reaches an expression containing only Green's functions with no external $h$ legs on the right hand side. 
The result of this repeated application of the recursion relation will therefore take the form
\begin{equation}
\label{schematicrecursion}
\Gamma_{N,M}=\sum_{\Delta m} a_{N,\Delta m} \Gamma_{0,M+\Delta m}.
\end{equation}
We proceed to calculate $a_{N,\Delta m}$.

We begin by labeling the term that lowers $N$ by 1 an $o$-type term and the term that lowers $N$ by 2 an $e$-type term. 
To keep track of different terms generated in the repeated application of \eqref{mastereq}, we construct strings of $e$'s and $o$'s. 
For example, $\lq \lq eoee"$ corresponds to a term generated by an e-type term in the first recursion followed by  an o-type term, followed by two e-type terms. 
Different strings with the same numbers of $e$'s and $o$'s correspond to terms 
with the same $\Delta m$ but with different intermediate coefficients.  
The coefficients of all possible strings of a given $\Delta m$ 
must be summed to give $a_{N,\Delta m}$. 

The absence of external $h$ legs on the right-hand side of eq. \eqref{schematicrecursion} forces
 $N=p+2q$ and $\Delta m = 2p +2q=2(N-q)$, 
 where $p$ is the number of $e$'s and $q$ is the number of $o$'s in a given string. 

For a given $N$, $q$ can take $\floor{\frac{N}{2}}+1$ different values since $q=0,1,\ldots, \floor{\frac{N}{2}}$, corresponding to $\floor{\frac{N}{2}}+1$ different values of $\Delta m$. 
Equation \eqref{schematicrecursion} therefore can  be rewritten as 
\begin{equation}
\label{schematicrecursionq}
\Gamma_{N,M}=\sum_{q=0}^{\floor{\frac{N}{2}}} a_{N,q} \Gamma_{0,M+2(N-q)}.
\end{equation}

For a given string of $e$'s and $o$'s, each $o$ contributes a factor of $\frac{\expval{H}}{M+1}$ and each $e$ contributes $\frac{N-1}{M+1}$. Clearly these depend on the position of the letter 
in the string, and the length of the string.


For a string of $e$'s and $o$'s, say $eooeoeo$, we label each $e$ and $o$ by a pair of indices $j,k$.
The leftmost entry in the string is labelled  $j=0,k=1$. 
Each $e$ or $o$ in the string increases $j$ by two (for the next entry); 
each $o$  increases $k$ by 1, while each $e$ increases $k$ by 2. 
The rightmost term for string of length $p+q$ will therefore be labelled by $j=2(N-q)$,
independent of the ordering of $e$'s and $o$'s,
while the $k$ label will depend on the ordering.
To each $o$ we can associate its correct factor (as given above), $\frac{\expval{H}}{M+1+j}$,
and to each $e$ its  factor, $\frac{N-k}{M+1+j}$.  
These are multiplied together to get the contribution of this string to $a_{N,q}$.

Disregarding momentarily the $N-k$ in the numerators of the $e$ factors, 
the rest of the contributions are the same for permutations of a string of $p$ $e$'s and $q$ $o$'s: 
\begin{equation}
\label{eqn:overallfactor}
\frac{\expval{H}^{(N-2q)}}{(M+1)(M+3) \ldots \big(M+2(N-q)+1\big)} = \frac{\expval{H}^{(N-2q)}(M+1){!}{!}}{(M+2N-2q-1){!}{!}}
\end{equation}

To find the correct numerator, we must construct all possible strings of $p$ $e$'s and $q$ $o$'s satisfying $p+2q = N$, evaluate their numerator, and then sum them.
For a given value of $p$ and $q$, there are therefore $\frac{(N-q){!}}{p{!}q{!}}$ different strings. 
Each $e$ in a string contributes $(N-k)$ to the numerator of that string. 
The full numerator 
(excluding the overall contribution included in equation \eqref{eqn:overallfactor}) 
can therefore be expressed as a nested sum:

\begin{equation}
\label{numeratorproof}
\sum_{i_{1}=1}^{N-(2q-1)}(N-i_{1})\Bigg(\sum_{i_{2}=i_{1}+2}^{N-(2q-3)}(N-i_{2}) 
\bigg(\ldots \Big(\sum_{i_{q-1}=i_{q-2}+2}^{N-3}(N-i_{q-1})  \big(\sum_{i_{q}=i_{q-1}+2}^{N-1}(N-i_{q})\big) \ldots \Bigg).
\end{equation}

The leftmost sum (over $i_1$) accounts for the possible locations of the leftmost $e$ in a string of length $(N-q)$. 
The second sum accounts for the possible locations of the second-leftmost $e$, and so on.
Since each  of the $q$  $e$'s in the string contributes a single sum, 
there are in total $q$ nested sums. 

To evaluate \eqref{numeratorproof}, we start by evaluating the rightmost (i.e. innermost) sum
\begin{equation}
\label{sum1}
\sum_{i_{q}=i_{q-1}+2}^{N-1}(N-i_{q})=\frac{1}{2}(N-i_{q-1}-1)(N-i_{q-1}-2).
\end{equation}
This can be written as
\begin{equation}
\label{eqn:firstsum}
\frac{1}{2}(N-i_{q-1}-2)_{2}\,,
\end{equation}
where $(x)_{n}$ is the Pochhammer symbol, defined by
\begin{equation}
\label{pochhammerdef}
(x)_{n} \equiv \frac{\Gamma(x+n)}{n} = n{!}\binom{x+n-1}{x-1}.
\end{equation}
A summation identity of Pochhammer symbols will prove very useful:
\begin{equation}
\label{pochhammersum}
\sum_{x=1}^{m}(x)_{n} = \frac{1}{n+1}(m)_{n+1}\,.
\end{equation}
This can be proved using the so-called multiset identity \cite{MRSpiegel}
\begin{equation}
\sum_{k=0}^{n}\binom{m+k-1}{k} = \binom{n+m}{n}.
\end{equation}
It is also easy to see that the Pochhammer symbol satisfies the recursion relation
\begin{equation}
\label{pochhammercomp}
(x+n)(x)_{n} = (x)_{n+1}.
\end{equation}

We can use equations \eqref{pochhammercomp} and \eqref{eqn:firstsum}
to express the second sum from the right in \eqref{numeratorproof} as
\begin{equation}
\sum_{i_{q-1}=i_{q-2}+2}^{N-3}\frac{1}{2}(N-i_{q-1}-2)_{3}.
\end{equation}
To use \eqref{pochhammersum}, we first rewrite this as
\begin{equation}
\sum_{i_{q-1}=1}^{N-i_{q-2}-4}\frac{1}{2}(N-i_{q-1}-i_{q-2}-3)_{3}\,.
\end{equation}
We then  define $x=N-i_{q-1}-i_{q-2}-3$ and  use \eqref{pochhammersum} to get
\begin{equation}
\sum_{x=1}^{N-i_{q-2}-4}\frac{1}{2}(x)_{3} = \frac{1}{8}(N-i_{q-2}-4)_{4}.
\end{equation}

We continue in a similar fashion 
through the rest of the $q$ nested sums in \eqref{numeratorproof}. 
The lower limit of each sum can be made 1 by shifting $i_{q-j}$ downward by $i_{q-j-1}+1$,
and then defining a dummy variable $x_j \equiv N-i_{q-j}-i_{q-j-1}-(2j+1)$. 
Each individual sum then takes the form
\begin{equation}
\label{nestedsumsindv}
\sum_{x_{j}=1}^{N-i_{q-(j+1)}-2(j+1)}\frac{1}{(2j){!}{!}}(x_j)_{2j-1}.
\end{equation} 

As we step outward (leftward) through the sums, 
the index $n$ of the Pochhammer symbol increases by 2, 
the argument of the Pochhammer symbol decreases by two, 
and we pick up an overall factor of  $\frac{1}{2j}$. 
We perform the first $q-1$ sums and we are left with
\begin{equation}
\label{numeratorcoef1}
\begin{split}
&\sum_{i_{1}=1}^{N-(2q-1)}\frac{1}{(2q-2){!}{!}}(N-i_{1})\big(N-i-(2q-2)\big)_{2q-2} \\ &= \sum_{i_{1}=1}^{N-(2q-1)}\frac{1}{(2q-2){!}{!}}\big(N-i-(2q-2)\big)_{2q-1} \\ &= \frac{N{!}}{(2q){!}{!}(N-2q){!}}
\end{split}
\end{equation}
where we have used the identities \eqref{pochhammercomp} and \eqref{pochhammersum}, 
and the definition of the Pochhammer symbol \eqref{pochhammerdef}.

Putting everything together, the coefficients $a_{N,q}$ become
\begin{equation}
\frac{\expval{H}^{N-2q}(M-1){!}{!}N{!}}{(2q){!}{!}\big(M+2(N-q)-1\big){!}{!}(N-2q){!}},
\end{equation}
and each $\Gamma_{N,M}$ is expressed as a linear superposition of $\Gamma_{0,m}$ as
\begin{equation}
\label{recursivesol}
\Gamma_{N,M} = \sum_{k=0}^{\floor{\frac{N}{2}}}\frac{(M-1)!! N!}{(M+2N-2k-1)!!} \frac{\expval{H}^{N-2k}}{(2k)!!(N-2k)!}\Gamma_{0,M+2(N-k)}. 
\end{equation}

The scalar-sector effective potential, appropriate for calculating processes containing only external scalars can be expressed generally as
\begin{equation}
\label{Ascalarveff}
V^{eff}_{\phi} = -\sum_{N,M=0}^{\infty}\frac{\Gamma_{N,M}}{N! M!}h^{N}\pi ^{M}.
\end{equation}

The LSS Theorem \eqref{adlersc} (i.e. the Adler self consistency condition for $N=0,M=2$)
ensures the masslessness of  $\pi$. 
The Adler self-consistency conditions \eqref{adlertmatrix}
also recursively force all Green's functions 
with an odd number of external $\pi$ legs to vanish. 
Eqns. \eqref{adlertmatrix} and \eqref{adlerscT} are enforced in equation \eqref{recursivesol} 
by defining $M=2m$ in \eqref{recursive}. Using \eqref{recursivesol} in \eqref{Ascalarveff} and
converting double factorials into normal ones and defining $j=N-q$, 
the scalar-sector effective potential becomes
\\~\\
\begin{equation}
\label{veffrev}
V^{eff}_{\phi} = -\sum_{m=0}^{\infty}\frac{\pi^{2m}}{m{!}}\sum_{N=0}^{\infty}h^{N}\sum_{j=\ceil{\frac{N}{2}}}^{N}\frac{\big(2\expval{H}\big)^{2j-N}}{(N-j){!}(2j-N){!}}\frac{(m+j){!}}{\big(2(m+j)\big){!}}\Gamma_{0,2(m+j)}.
\end{equation}

For a fixed value of $m$, because of the sum on $N$, 
the index $j$ runs from $0$ to $\infty$. 
We investigate the coefficients of $\Gamma_{0,2(m+j)}$ 
for a fixed value of $j$ and note that $N$ has to run from $j$ to $2j$. 
If we define $N=i+j$, the range of $N$ forces $i$ to run from $0$ to $j$. 
If we also define $n=m+j$, then the r.h.s. of \eqref{veffrev} can be rewritten more simply as
\begin{equation}
\sum_{m=0}^{\infty}\frac{\pi^{2m}}{m{!}}
\sum_{n=m}^{\infty}\frac{\Gamma_{0,2n}n{!}}{(2n){!}}
\sum_{i=0}^{n-m}\frac{h^{2i}\big(2\expval{H}h\big)^{n-i-m}}{i{!}(n-i-m){!}}.
\end{equation}
To find the coefficients of $\Gamma_{0,2n}$ for a fixed value of $n$, 
where $n$ ranges from 0 to $\infty$, note that, because of the sum on $n$ from $m$ to $\infty$, 
the terms that contain $\Gamma_{0,2n}$ for a fixed $n$ 
can only come from $m=0,1,\ldots ,n$. 
After relabelling $m=j$, this observation leads us 
to the following expression for the r.h.s of \eqref{veffrev}
\begin{equation}
\label{vefftrin}
\sum_{n=0}^{\infty}\frac{\Gamma_{0,2n}n{!}}{(2n){!}}
\sum_{j=0}^{n}\frac{\pi^{2j}}{j{!}}
\sum_{i=0}^{n-j}\frac{h^{2i}\big(2\expval{H}h\big)^{n-i-j}}{i{!}(n-i-j){!}}.
\end{equation}
The series in \eqref{vefftrin} contains a trinomial expansion
\begin{equation}
n{!}\sum_{j=0}^{n}\frac{\pi^{2j}}{j{!}}
\sum_{i=0}^{n-j}\frac{h^{2i}\big(2\expval{H}h\big)^{n-i-j}}{i{!}(n-i-j){!}}  =\Big(h^{2}+\pi^{2}+2\expval{H}h\Big)^{n}\nonumber
\end{equation}
and thus the effective potential is expressed as in \eqref{veffxi}.

\end{document}